\documentclass[a4paper,fleqn]{cas-dc}

\usepackage[numbers]{natbib}

\def\tsc#1{\csdef{#1}{\textsc{\lowercase{#1}}\xspace}}
\tsc{WGM}
\tsc{QE}

\begin{document}
\let\WriteBookmarks\relax
\def\floatpagepagefraction{1}
\def\textpagefraction{.001}

\title [mode = title]{A Generalizable TCAD Framework for Silicon FinFET Spin Qubit Devices with Electrical Control}
\shorttitle{TCAD Framework for Silicon FinFET Spin Qubit Devices}
\shortauthors{Q. Ding et al.}

\author[1]{Qian Ding}
\cormark[1]
\ead{dingq@iis.ee.ethz.ch}
\affiliation[1]{organization={ETH Zurich, Integrated Systems Laboratory},
                addressline={Gloriastrasse 35},
                city={Zurich},
                postcode={8092},
                country={Switzerland}}

\author[2]{Andreas V. Kuhlmann}
\affiliation[2]{organization={University of Basel, Department of Physics},
		addressline={Klingelbergstrasse 82},
		city={Basel},
                postcode={4056},
                country={Switzerland}}

\author[3]{Andreas Fuhrer}
\affiliation[3]{organization={IBM Research Europe-Zurich, Quantum Technology \& Computing},
                addressline={S\"aumerstrasse 4},
                city={R\"uschlikon},
                postcode={8803},
                country={Switzerland}}

\author[1]{Andreas Schenk}

\cortext[cor1]{Corresponding author. E-mail address:}

\begin{abstract}[S U M M A R Y]
We present a TCAD-based simulation framework established for quantum dot spin qubits in a silicon FinFET platform with all-electrical control of the spin state. The framework works down to 1\,K and consists of a two-step simulation chain, from definition of the quantum dot confinement potential with DC bias voltages, to calculation of microwave response electric field at qubit locations using small-signal AC analysis. An average field polarization vector at each quantum dot is extracted via a post-processing step. We demonstrate functionality of this approach by simulation of a recently reported two-qubit device in the form of a 5-gate silicon FinFET. The impact of the number of holes in each quantum dot on the MW response $E$-field polarization direction is further investigated for this device. The framework is easily generalizable to study future multi-qubit large-scale systems.
\end{abstract}


\maketitle


\section{Introduction}\label{intro}

\vspace{0.2cm}
Scalability is vital for building useful quantum computers with quantum error correction, but a tough task with respect to actual physical implementation. One promising platform to overcome this challenge is quantum dot (QD) spin qubits embedded in multi-gate silicon FinFETs \cite{loss98,golovach06}. Recently, hole spin qubits hosted by double QDs in a 5-gate silicon FinFET that can operate above 4\,K have been reported \cite{nat}. The device fabrication is compatible with standard CMOS technology \cite{aip}, and qubit manipulation is realized by electric dipole spin resonance (EDSR) with microwave (MW) electrical signals applied to a single gate electrode. This makes it a good candidate towards large-scale integration of spin qubit devices. To scale up the system in the near future, a simulation-aided analysis for the design of all-electrical qubit control is highly desirable. For this purpose, we developed a TCAD-based framework that can perform DC and AC simulations down to 1\,K. The MW response electric field ($E$-field) polarization vector averaged over each QD is extracted in post-processing steps. Gate cross-talk is also included in these AC simulations by a first-order capacitive coupling model.

\begin{figure}[htbp]
\centerline{\includegraphics[width=0.45\textwidth, height=9cm]{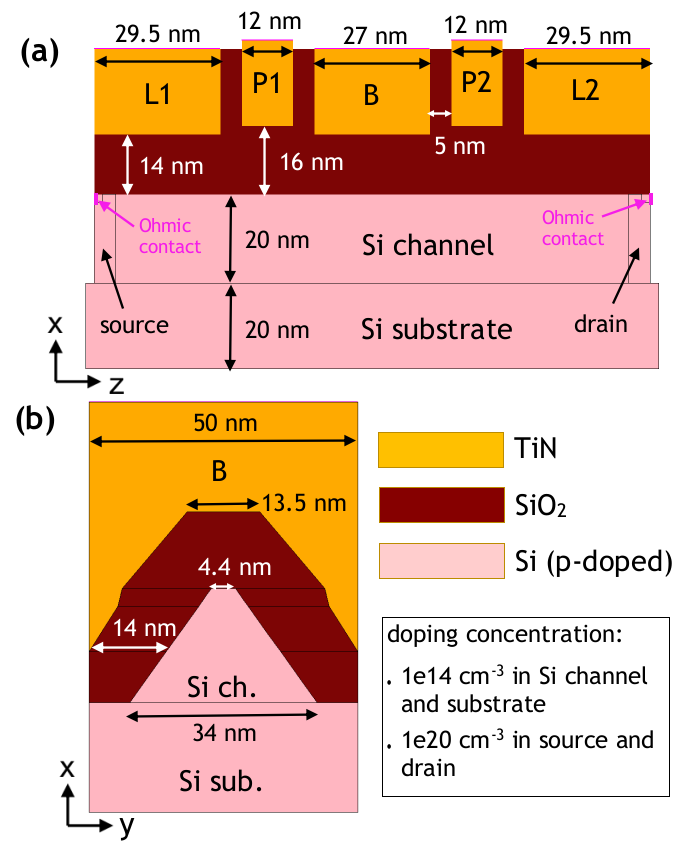}}
\caption{Sketch of simulated 5-gate Si FinFET. (a)/(b) side/cross section view along/vertical to fin direction. MW signal (5 GHz, 12 mV amplitude) for spin qubit control is applied on gate P1. 
}
\label{fig1}
\end{figure}

\vspace{0.2cm}
We illustrate the simulation framework by taking the reported two-qubit device \cite{nat} as an example, while the generalization to multi-qubit devices is straightforward. 
The simulated device structure is shown in Fig.~\ref{fig1}.
In the following, we first introduce the simulation workflow and explain the AC method for MW $E$-field polarization vector calculation, including the gate cross-talk estimation model. Then, we employ an example of single-hole QDs to show how AC simulation and post-processing work. Afterwards, we discuss the impact of the number of holes in the QDs on the averaged response $E$-field polarization vector.

\section{Simulation Methodology}\label{simu}

\begin{figure}[htbp]
\centerline{\includegraphics[width=0.47\textwidth, height=7.2cm]{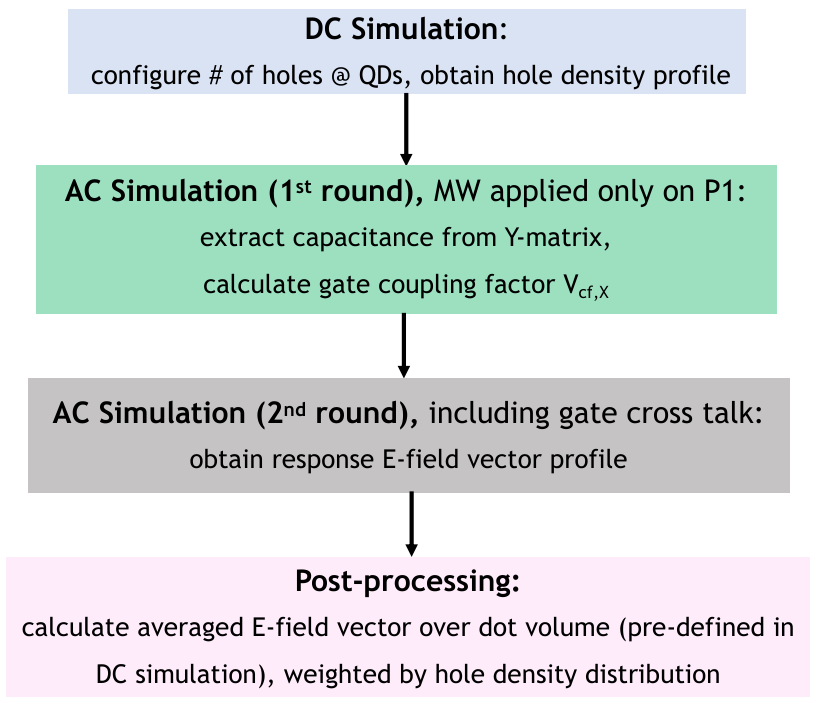}}
\caption{Simulation workflow for calculation of MW response $E$-field polarization vector.}
\label{fig2}
\end{figure}

The entire simulation chain includes four steps, as shown in Fig.~\ref{fig2}.
To calculate the response $E$-field polarization vectors, first, we run a quasi-stationary DC simulation to generate QDs with a specific number of holes. 
Quantum confinement is modeled with the density-gradient method \cite{SDev,wett01}, where a potential-like quantity $\Gamma_{p}$ is derived solving an additional equation. The hole density is then obtained from
	\begin{equation}
		p({\bf r}) = N_\mathrm{v}F_{1/2}\Big[\beta\big(E_\mathrm{v}({\bf r}) + \Gamma_\mathrm{p}({\bf r}) - E_\mathrm{f,p}\big)\Big] \label{pDens}
	\end{equation}
with $\beta = 1/k_\mathrm{B}T$ and $E_\mathrm{f,p}=0$. Obviously, Eq.~(\ref{pDens}) is based on local thermodynamic equilibrium and Fermi-Dirac statistics, which breaks down in the SET regime of the transistor. Options like Gibbs statistics are not available in S-Device. As a consequence, the dot charge changes continuously with gate voltage, and no tunnel barriers can be generated in the limit $T\rightarrow{}0$\,K. Transverse confinement is in good agreement with 2D k$\cdot{}$p Schr\"odinger-Poisson reference calculations \cite{SDev}, but longitudinal confinement effects cannot be easily calibrated. Therefore, the exact density overlap between the dots remains vague. However, this is not expected to impact the AC analysis significantly.

In the second step, AC simulations are performed using the electrical small-signal analysis method \cite{SDev}. 
This method is valid for qubit device simulation because the MW signal amplitude is usually much smaller than the applied DC bias and the device size (few hundreds of nanometers) is much smaller than the MW wavelength (centimetre range). 

\begin{figure}[htbp]
\centerline{\includegraphics[width=0.5\textwidth, height=7cm]{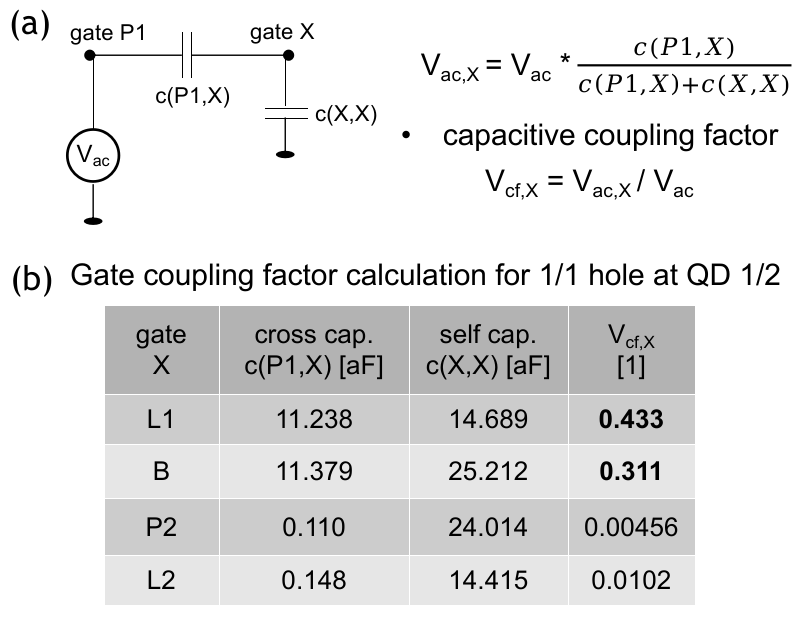}}
\caption{(a) Simplified 1\textsuperscript{st}-order capacitive coupling circuit model for gate cross talk calculation. (b) Calculated voltage coupling factor for the case of two single-hole QDs.}
\label{fig3}
\end{figure}

To take the cross talk between gates into account, we introduce a simplified capacitive coupling model based on the first-order approximation. This is achieved by running two AC simulation rounds.
A first round is performed with the AC signal applied on gate P1 only to extract the Y-matrix of the device.
The obtained capacitance elements are then used to calculate a capacitive coupling factor V\textsubscript{cf,X} between P1 and any other gate X, based on a voltage divider circuit (see Fig.~\ref{fig3} (a)).
Then, a second-round simulation is performed with AC signals also applied to other gates, where their AC voltage amplitudes depend on their respective coupling factors V\textsubscript{cf,X}. Fig.~\ref{fig3} (b) shows the results of the coupling factors in case of one hole in each QD. The coupling is strong only for gates L1 and B that are close to P1.

\vspace{0.2cm}
The desired AC-response $E$-field polarization vector is obtained based on the S-Device default output $\Im(J_D)$ after the second AC run including the gate cross talk. 
This extraction relies on the following relations:
\begin{equation}
        E = -\nabla\varphi \label{Efield}
\end{equation}
\vspace{-0.4cm}
\begin{equation}
        J_D = -i\omega\epsilon\nabla\varphi \label{dispCur}
\end{equation}
\vspace{-0.4cm}
\begin{equation}
        \Re(E) = \Im(J_D)/\omega\epsilon \label{RealE}
\end{equation}
According to Eq.~(\ref{RealE}), the imaginary part of the displacement current response $\Im(J_D)$ (default output) is representative for the real $E$-field response $\Re(E)$. Their magnitudes differ only by a scaling factor, whereas the vector directions are exactly the same. This one-to-one correspondence facilitates the subsequent calculation of a normalized field polarization vector averaged over the QD for each qubit by a post-processing step (see the results in Sec.~\ref{SimuRe}).

\section{Simulation Results}\label{SimuRe}


In this section, we first present results obtained for a (1,1) charge configuration with one hole in each QD, to demonstrate the simulation workflow. Then we study the influence of an increasing hole number on the field polarization, as this parameter can be hard to determine experimentally.

\subsection{AC-response field for single-hole QDs}\label{OneDotResp}
\begin{figure}[htbp]
\centerline{\includegraphics[width=0.5\textwidth, height=11.8cm]{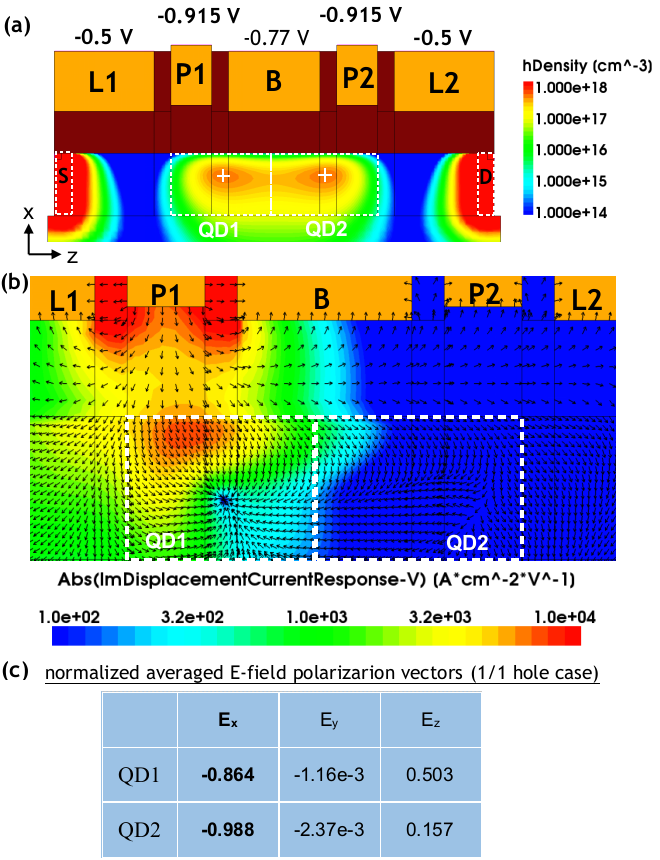}}
\caption{
	(a) DC hole density profile along the fin direction. White dashed lines (cross labels) indicate the QD volume used for integration to obtain the number of holes (dot centers).
(b) $\Im(J_D)$ response profile, taken at V\textsubscript{ac} = -12 mV. The arrows exactly represent the field polarization.
(c) Calculated normalized x/y/z-components of the averaged response field polarization vector at each QD.
}
\label{fig4}
\end{figure}

\vspace{0.2cm}
Fig.~\ref{fig4} (a) shows the hole density profile in presence of two single-hole QDs, obtained from the DC simulation. The QD hole number is calculated by integrating the hole density over a defined quantum dot volume (indicated by white dashed lines).  Then, after running two-round AC simulations, the field response vector profile is calculated (see Fig.~\ref{fig4} (b)). Two singularities show up in the AC-response $E$-field vector distribution due to the low response at the dot centers ((labeled by white crosses in Fig.~\ref{fig4} (a)).
In order to assign a single field polarization vector to the three-dimensional distribution, we introduce a normalized field vector averaged over the dot volume weighted by the DC hole density. It is calculated in post-processing by multiplying the hole density with each of the x/y/z-components of the field vector, then integrating the resulting quantity over the dot volume, and finally dividing the integrals by their root sum square. In this way the DC hole density acts as a weighting factor in the field extraction procedure.

The calculated normalized average field polarization vectors for the QDs in a (1,1) charge configuration are shown in Fig.~\ref{fig4} (c).
As we see, the $E$-field vectors at both QDs are mostly polarized along the -x-direction. This observation is related to the specific location of the QD centers. 
For QD1, as its center is shifted towards gate B, a hotspot of the response field amplitude occurs directly in the dot region under gate P1 (see Fig.~\ref{fig4} (b)). 
This results in a larger contribution pointing along the -x axis when averaging the field vectors over the QD volume.
For QD2, there is no clear hotspot, but the vertical (-x-direction) components of the field vectors in the upper part of the QD contribute more.  

\subsection{Impact of number of holes at QDs}\label{HoleNum}
\begin{figure}[htbp]
\centerline{\includegraphics[width=0.48\textwidth, height=7.8cm]{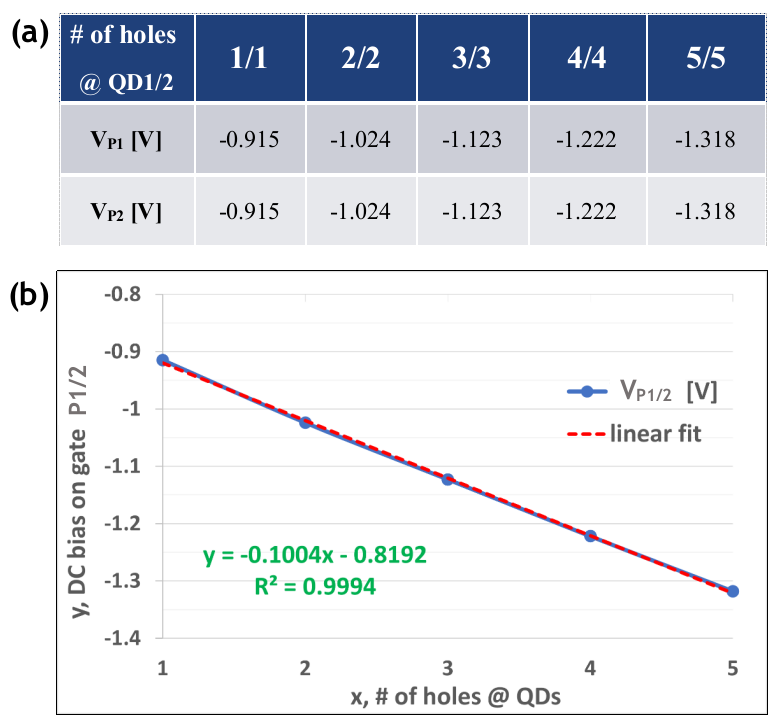}}
\caption{(a) Values of DC bias voltages on plunger gates P1/P2 to create hole configurations from (1,1) to (5,5) in (QD1,QD2). (b) DC bias on plunger gates vs. number of holes at QDs. The linear fit (red dashed line) is given by the green-colored equation. The slope represents the required change of DC bias on gate P1/2 to accumulate one more hole at the QDs.
}
\label{fig5}
\end{figure}

To tune the hole number simultaneously at both QDs, we choose to adjust the DC bias on the plunger gates P1 and P2, while keeping the bias values on all the other gates unchanged. The required plunger gate voltages for inducing hole configurations from (1,1) to (5,5) in (QD1,QD2) are shown in Fig.~\ref{fig5} (a).
It turns out that the relation between P-gate voltage and QD hole number is almost linear, as seen from the fitted line (red dashed) in Fig.~\ref{fig5} (b). From the extracted slope one can infer that, in order to accumulate one more hole on each QD, the gate bias on P1 and P2 should be reduced simultaneously by $\sim$ 0.1 V. This observation could be useful for future device design to improve qubit control.

\begin{figure}[htbp]
\centerline{\includegraphics[width=0.5\textwidth, height=11.4cm]{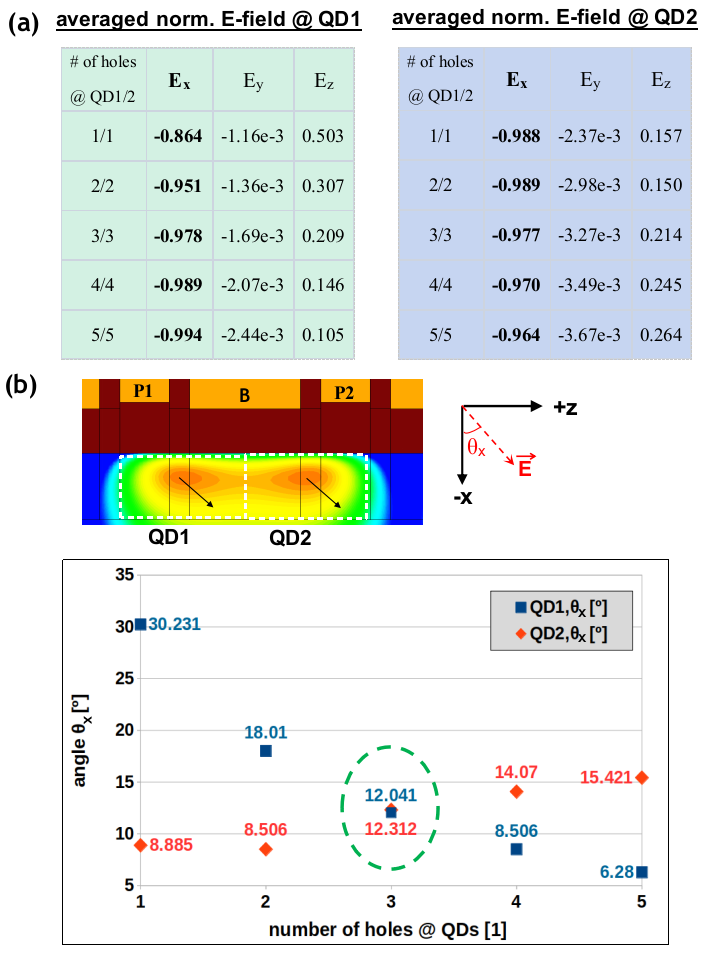}}
\caption{(a) Calculated normalized x/y/z-components of the averaged response-field polarization vector at QD1/2 for hole configurations from (1,1) to (5,5) at (QD1,QD2). (b) Plot of response field polarization angle $\theta_x$ vs. number of holes. The slope represents the required change of DC bias on P1/2-gates to accumulate another hole at the QDs.
}
\label{fig6}
\end{figure}

\vspace{0.2cm}
The impact of an increasing hole number on the averaged field polarization vector of the QDs is shown in Fig.~\ref{fig6} (a).
From the highlighted values in the tables we conclude that as the dot hosts more holes, the field polarization along -x-direction becomes stronger/weaker at QD1/QD2 respectively. However, the size of this effect is much smaller for QD2 than for QD1 simply because of the longer distance to the AC control gate P1. As a further post-processing step, instead of directly using the normalized field component, we define a field polarization angle, which is the angle between the extracted $E$-field vector and the -x-axis (see the upper plot in Fig.~\ref{fig6} (b)). The dependence of this field polarization angle on the number of holes (see the lower plot in Fig.~\ref{fig6} (b)) gradually saturates with increasing hole number.  A special situation occurs for the (3,3) hole configuration, where the polarization angles become almost the same at both QDs, as a consequence of the opposite trends (slopes) for QD1 and QD2.

\section{Conclusion}
A TCAD-based simulation framework for the computation of the microwave response $E$-field polarization is demonstrated using a 5-gate FinFET hole spin qubit device. The extracted field polarization angle at the qubit location will be used for future study of the Rabi driving strength. We showed that the location of the center of the quantum dot has a strong influence on the average field polarization. The latter also depends on the number of holes in the QDs, but this effect quickly saturates with increasing hole number. 

\section*{Acknowledgments}
The authors thank Oleg Penzin and Paul Pf\"affli from Synopsys Inc. for their valuable suggestions and practical help to make Sentaurus-Device a successful tool in this challenging project. This work was partially supported by the NCCR SPIN.

\bibliographystyle{cas-model2-names}

\nocite{*}%
\bibliographystyle{elsarticle-num}

\bio{auBio/QD}
Qian Ding is PhD student at the Integrated Systems Laboratory (IIS) of ETH Zurich. Her research focus is on the simulation and design of nanophotonic and FinFET spin qubit devices in collaboration with IBM-Research Zurich and the University of Basel. 
\endbio

\bio{auBio/AK}
Andreas Kuhlmann received his Ph. D. degree in physics from University of Basel in 2013. After a postdoc at IBM Research Europe-Zurich, he rejoined the University of Basel as a Georg H. Endress Fellow in 2018 and became a senior scientist in 2021. His research focus lies on quantum computing with hole spin qubits integrated in silicon FinFET devices. He has co-authored over 25 peer reviewed papers.
\endbio

\bio{auBio/AF}
Andreas Fuhrer received his Ph.D. degree in physics from the Swiss Federal Institute of Technology in Zürich (ETHZ) in 2003. After postdocs with Prof. L. Samuelson, Sweden and Prof. M.Y. Simmons, Australia, he joined IBM Research Europe - Zurich in 2008.  His research interests include scanning probe based fabrication, spintronics and quantum computing with both superconducting qubits and spin qubits in semiconductor quantum dots. Since 2014 he is specifically interested in hole spin qubits defined in transistor like bulk finFET devices. He has co-authored over 60 peer reviewed papers and holds 9 patents.
\endbio

\bio{auBio/AS}
Prof. Dr. Andreas Schenk had been heading the Nano-Device Physics Group at the Integrated Systems Laboratory (IIS) of ETH Zurich. Since 1991 he had been working as scientific adjoint at the IIS, where he habilitated in 1997 and became honorary professor in 2004. His research focus is on the physics-based modeling of nano- and optoelectronic devices. He authored and co-authored two books and more than 200 refereed papers.
\endbio

\end{document}